\documentclass[journal]{IEEEtran}

\ifCLASSINFOpdf
\else
\fi
\usepackage{amsfonts}
\usepackage{cite}
\usepackage{hyperref} 
\hypersetup{hidelinks}  
\usepackage{amsmath} 
\usepackage{graphicx}
\usepackage{algorithmic, algorithm}
\usepackage{color}
\usepackage{multirow}

\begin{document}


\title{Preventive-Corrective Cyber-Defense: \\Attack-Induced Region Minimization and Cybersecurity Margin Maximization}

\author{Jiazuo~Hou,~\IEEEmembership{Graduate~Student~Member,~IEEE,}
Fei~Teng,~\IEEEmembership{Senior~Member,~IEEE}
Wenqian~Yin,~\IEEEmembership{Graduate~Student~Member,~IEEE,}
Yue~Song,~\IEEEmembership{Member,~IEEE,}
and~Yunhe~Hou,~\IEEEmembership{Senior~Member,~IEEE}
\thanks{This work was supported in part by
the National Natural Science Foundation of China (NSFC) under Grant 52177118,
in part by the Collaborative Research Fund (CRF Project No. C1052-21GF) from the Research Grants Council, Hong Kong SAR,
and in part by the Fundamental Research Funds for the Central Universities under Grant 22120230432.}
\thanks{Jiazuo Hou, Wenqian Yin, and Yunhe Hou are with the Department of Electrical and Electronic Engineering, The University of Hong Kong, Hong Kong SAR; and are also with The University of Hong Kong Shenzhen Institute of Research and Innovation, Shenzhen 518057, China (e-mail: houjz@nus.edu.sg, wqyin@eee.hku.hk, yhhou@eee.hku.hk).}
\thanks{Fei Teng is with the Department of Electrical and Electronic Engineering, Imperial College London, SW7 2AZ London, U.K. (e-mail: f.teng@imperial.ac.uk).}
\thanks{Yue Song is with the Department of Control Science and Engineering, Tongji University, Shanghai 201804, China, also with the National Key Laboratory of Autonomous Intelligent Unmanned Systems, Shanghai 201210, China, and also with the Frontiers Science Center for Intelligent Autonomous Systems, Ministry of Education, Shanghai 200120, China (e-mail: ysong@tongji.edu.cn).}
\vspace{-20pt}}




\maketitle

\begin{abstract}
False data injection (FDI) cyber-attacks on power systems can be prevented by strategically selecting and protecting a sufficiently large measurement subset, which, however, requires adequate cyber-defense resources for measurement protection. With any given cyber-defense resource, this paper proposes a preventive-corrective cyber-defense strategy, which minimizes the FDI attack-induced region in a preventive manner, followed by maximizing the cybersecurity margin in a corrective manner. First, this paper proposes a preventive cyber-defense strategy that minimizes the volume of the FDI attack-induced region via preventive allocation of any given measurement protection resource. Particularly, a sufficient condition for constructing the FDI unattackable lines is proposed, indicating that the FDI cyber-attack could be locally rather than globally prevented. Then, given a non-empty FDI attack-induced region, this paper proposes a corrective cyber-defense strategy that maximizes the cybersecurity margin, leading to a trade-off between the safest-but-expensive operation point (i.e., Euclidean Chebyshev center) and the cheapest-but-dangerous operation point. Simulation results on a modified IEEE 14 bus system verify the effectiveness and cost-effectiveness of the proposed preventive-corrective cyber-defense strategy.
\end{abstract}

\begin{IEEEkeywords}
Attack-Induced region,
cybersecurity margin,
Euclidean Chebyshev center,
insufficient cyber-defense resource,
preventive-corrective cyber-defense.
\end{IEEEkeywords}


\section*{Nomenclature}
\addcontentsline{toc}{section}{Nomenclature}
\begin{IEEEdescription}[\IEEEusemathlabelsep\IEEEsetlabelwidth{$V_1,V_2,V_3,V_3$}]
\item[System Variable]
\item[$\boldsymbol{D}$, $\boldsymbol{G}$, $\boldsymbol{F}$] load, generation, and line flow vector
\item[$\underline{\boldsymbol{G}}$, $\overline{\boldsymbol{G}}$] lower and upper limits of generation vector
\item[$\underline{\boldsymbol{F}}$, $\overline{\boldsymbol{F}}$] lower and upper limits of line flow vector
\item[$\boldsymbol{S}$] shifting factor matrix, whose $n$-th row vector is $\boldsymbol{S}_n$
\item[$\boldsymbol{U_G}$, $\boldsymbol{U_D}$] bus-generator incidence matrix and bus-load incidence matrix
\item[$\mathcal{D}$, $\mathcal{L}$] set of all loads and all lines
\item[$\boldsymbol{c}$] generation cost vector
\item[$\Theta^{\text{security}}$] security region
\item[$\Theta^{\text{insecurity}}$] insecurity region

\item[]
\item[Cyber-Attack Variable]
\item[$\Delta \boldsymbol{D} = \{ \Delta D_d  \} $] FDI attack injection on the $d$-th load measurement
\item[$\Delta \boldsymbol{F} = \{ \Delta F_n  \}$] FDI attack injection on the $n$-th line flow measurement
\item[$\Delta \boldsymbol{D}^n$] FDI attack injection vector with respect to the $n$-th line
\item[$\boldsymbol{\tau} = \{ \tau_d \}$] FDI attacking ability on the $d$-th load measurement
\item[$\boldsymbol{H},~\boldsymbol{V}$] maximum line overloading in two opposite line flow directions
\item[$\Omega^{\text{ARR}}$] FDI attack-reachable region (ARR)
\item[$\Omega^{\text{AIR}}$] FDI attack-induced region (AIR)
\item[$\Gamma^{\text{security}}$] preventive security region

\item[]
\item[Cyber-Defense Variable]
\item[$\boldsymbol{\delta} = \{ \delta_d \}$] cyber-defense meter placement state (binary variable) of the $d$-th load measurement
\item[$\boldsymbol{\varepsilon} = \{ \varepsilon_n \}$] cyber-defense meter placement state (binary variable) of the $n$-th line flow measurement
\item[$B^{\text{AIR}}$, $C^{\text{AIR}}$] cyber-defense benefit and cost of the proposed shaping cyber-defense strategy
\item[$\omega^{\text{AIR}}$, $\overline{C^{\text{AIR}}}$] cyber-defense cost coefficient and cost budget of the proposed shaping cyber-defense strategy
\item[$B^G$, $C^G$] cyber-defense benefit and cost of the proposed dispatching cyber-defense strategy
\item[$\omega^G$] cyber-defense cost coefficient of the proposed dispatching cyber-defense strategy
\item[$r(\boldsymbol{G})$] cybersecurity margin of operation point $\boldsymbol{G}$
\item[$\lambda^n$] Lagrange multiplier associated with the power balance equation of the $n$-th line-oriented FDI attack
\item[$\underline{\boldsymbol{\alpha}^n}$, $\overline{\boldsymbol{\alpha}^n}$] Lagrange multiplier vectors associated with the load injection limits of the $n$-th line-oriented FDI attack
\item[$\underline{\boldsymbol{\beta}^n}$, $\overline{\boldsymbol{\beta}^n}$] Lagrange multiplier vectors associated with the line injection limits of the $n$-th line-oriented FDI attack
\item[$\boldsymbol{u}_{\underline{\boldsymbol{\alpha}^n}}$, $\boldsymbol{u}_{\overline{\boldsymbol{\alpha}^n}}$] additional binary vector regarding the complementary slackness conditions of $\underline{\boldsymbol{\alpha}^n}$ and $\overline{\boldsymbol{\alpha}^n}$
\item[$\boldsymbol{u}_{\underline{\boldsymbol{\beta}^n}}$, $\boldsymbol{u}_{\overline{\boldsymbol{\beta}^n}}$] additional binary vector regarding the complementary slackness conditions of $\underline{\boldsymbol{\beta}^n}$ and $\overline{\boldsymbol{\beta}^n}$
\item[$M$, $N$, $K$] sufficiently large positive constants
\end{IEEEdescription}

\vspace{5pt}
\section{Introduction}
\vspace{5pt}

\IEEEPARstart{R}{ecent} years have witnessed increasing reports on cyber-attack events~\cite{tech_IEA2020Cyber}, some of which result in severe damage to power systems~\cite{Ukraine_TPS2017}.
Therein, the false data injection (FDI) cyber-attack~\cite{liu2011false} is widely investigated in power systems~\cite{FDI_survey_TSG_2020} due to its stealthiness, i.e., the ability to compromise measurement data while bypassing the bad data detection (e.g., the largest normalized residual test~\cite{Abur_book_2004}).
The FDI cyber-attack has been demonstrated to achieve various attacking purposes, e.g., the economic loss~\cite{FDI_market_TSG2019}, 
the stability deterioration~\cite{SSAS_TSG2023}, and more importantly, the line overloading that could lead to not only instant line tripping~\cite{FDIOverLoad_TSG2019} but also subsequent cascading failure and even blackout~\cite{LRload_2012}.

To prevent the stealthy FDI cyber-attacks and mitigate attack-induced consequences, the \textit{preventive} cyber-defense strategies~\cite{bobba2010detecting} in system planning and \textit{corrective} cyber-defense strategies~\cite{TSG2023_AGCCorrect} in system operation are investigated, respectively.
One of the widely-applied preventive countermeasures is to
strategically select and protect a sufficiently large subset of measurement such that no stealthy FDI cyber-attack can be launched.
The pioneering work~\cite{bobba2010detecting} investigates how to select an optimal measurement subset to independently verify a few chosen state variables.
Many efforts, e.g., graph theory-based approaches~\cite{TII2021_GraphSensor, TSG2014_GraphSensor}, game theory-based methods~\cite{TSG2013_Game_Sensor}, are made to identify the minimum dimension and corresponding allocation of the measurement subset.
Moreover, fast greedy algorithms~\cite{TII2015_Greedy_Sensor} are proposed to quickly find the minimum measurement subset by solving combinatorial optimization problems~\cite{TSG2019_Comb_Sensor}.

The previous preventive countermeasures~\cite{bobba2010detecting, TII2021_GraphSensor, TSG2014_GraphSensor, TSG2013_Game_Sensor, TII2015_Greedy_Sensor, TSG2019_Comb_Sensor} focus on whether the selected measurement subset fails (denoted as $0$) or manages (denoted as $1$) to completely prevent any stealthy FDI cyber-attack, indicating a binary cyber-defense solution, i.e., $\{ 0,1 \}$.
This relies on a strong assumption that the cyber-defense resources should be sufficient to protect all the measurement in the selected subset.
Nevertheless, few efforts are made to investigate the scenario when the measurement protection is insufficient~\cite{ENTSO_2020} such that
not all the stealthy FDI cyber-attacks can be avoided,
indicating degraded cyber-defense solutions within $[0, 1]$.
This yields a non-empty \textit{FDI attack-induced region},
which threatens the power systems and thus needs to be minimized via the optimal allocation of insufficient measurement protection in a preventive manner.

Given that the stealthy FDI cyber-attacks may not be completely prevented due to insufficient measurement protection~\cite{InsuranceModel_Lau_TSG2020}, the most economic operation point, which usually approaches line flow limits, needs to be re-dispatch in a corrective manner to avoid attack-induced line overloading.
Specifically, to avoid both the cheapest-but-dangerous operation point and the safest-but-expensive operation point, the corrective re-dispatch should achieve a trade-off between the additional defense-induced operation cost and the \textit{cybersecurity margin}.
Similar to the physical security margin defined in the physical security region~\cite{Lee_TCNS2019}, the cybersecurity margin in this paper indicates 
the minimum distance from the operation point to the boundaries regarding the FDI attack-induced region.

Combining both pre-event preventive strategies and during-event corrective strategies, the two-stage preventive-corrective strategies are well studied to enhance the power system \textit{physical} resilience, including critical load restoration~\cite{TSG2018_ShunboLei}, transient stability maintenance~\cite{TPS2020_PreCor}, etc.
However, few efforts concentrate on the preventive-corrective strategies to enhance the power system \textit{cyber} resilience against cyber-attack events.
Then, intriguing questions arise:
\begin{itemize}
    \item \textit{Preventive cyber-defense strategy in system planning}:
    How to quantify and minimize the FDI attack-induced region via the preventive allocation of insufficient measurement protection?
    \item \textit{Corrective cyber-defense strategy in system operation}:
    Given the FDI attack-induced region from imperfect preventive countermeasures,
    how to quantify and enlarge the cybersecurity margin via the corrective re-dispatch of operation points?
    \item \textit{Cost-Benefit Pareto optimal front}:
    How to maximize cyber-defense benefits while minimizing cyber-defense costs, leading to Pareto optimal fronts in bi-objective optimization problems?
\end{itemize}

Motivated by the above questions, this paper proposes a cost-effective preventive-corrective cyber-defense strategy, which minimizes the FDI attack-induced region in a preventive manner, followed by maximizing the cybersecurity margin in a corrective manner.
The contributions of this paper are summarized as follows.
\begin{enumerate}
    \item This paper proposes a preventive cyber-defense strategy to characterize the FDI attack-induced region, whose volume is optimally shaped via preventive allocation of any given cyber-defense resource.
    In particular, a sufficient condition for constructing the FDI unattackable lines is proposed, indicating that the FDI cyber-attack could be locally rather than globally prevented.
    \item The proposed preventive cyber-defense strategy is mathematically formulated as a bi-objective bi-level (one-leader-multi-follower) problem, which is converted into a single-level mixed-integer linear programming (MILP) problem.
    Moreover, the closed-form lower bounds for three big-M constants are investigated.
    \item 
    Given a non-empty FDI attack-induced region, this paper proposes a corrective cyber-defense strategy to achieve a trade-off between maximizing the cybersecurity margin and minimizing the additional defense-induced operation cost.
    This leads to a balance between the safest-but-expensive operation point (i.e., Euclidean Chebyshev center) and the cheapest-but-dangerous operation point.
\end{enumerate}

The remaining of this paper is organized as follows.
Section~II illustrates the preliminaries of the FDI cyber-attack.
To limit the FDI attack-induced region,
Section~III proposes a preventive cyber-defense strategy,
which is reformulated into a single-level MILP problem in Section~IV.
Section~V proposes a corrective cyber-defense strategy to maximize the cybersecurity margin in a preventive security region.
Section~VI verifies the effectiveness and cost-effectiveness of the proposed preventive-corrective cyber-defense strategy.
Section~VII concludes this paper.

\section{Preliminaries of The False Data Injection Cyber-Attack} \label{se_preli}

Consider a general optimal power flow (OPF) problem that is subjected to the power balance~\eqref{eq_powerbalance}, the generation limits~\eqref{eq_Genlimit}, and the power flow equation and limits~\eqref{eq_Linelimit}:
\begin{alignat}{1}
    &\boldsymbol{1}^T \boldsymbol{D} = \boldsymbol{1}^T \boldsymbol{G} \label{eq_powerbalance} \\ 
    &\underline{\boldsymbol{G}} \leq \boldsymbol{G} \leq \overline{\boldsymbol{G}} \label{eq_Genlimit} \\ 
    &-\overline{\boldsymbol{F}} \leq \boldsymbol{F} = \boldsymbol{S} \, (\boldsymbol{U_G} \, \boldsymbol{G} - \boldsymbol{U_D} \, \boldsymbol{D}) \leq \overline{\boldsymbol{F}} \label{eq_Linelimit}
\end{alignat}
where $\boldsymbol{F}$, $\boldsymbol{D}$, and $\boldsymbol{G}$ are the line flow vector, the load vector, and the generation vector, respectively.
$\boldsymbol{S}$ is the shifting factor matrix.
$\boldsymbol{U_G}$ and $\boldsymbol{U_D}$ are the bus-generator and bus-load incidence matrix, respectively.
$T$ represents transpose operation.
$\overline{\boldsymbol{F}} > 0$ represents the line upper limits.
$\overline{\boldsymbol{G}} > 0$ and $\underline{\boldsymbol{G}} \ge 0$ represent the generator upper and lower limits, respectively.

Assume the generation measurement is not easy to be attacked~\cite{LR_2011}, leading to $\Delta \boldsymbol{G} = 0$, then we have the widely demonstrated fact.

\vspace{2pt}
\noindent
\textit{Fact 1. (Stealthiness of FDI Cyber-Attack)}
The FDI cyber-attack~\cite{liu2011false} is able to bypass the bad data detection (i.e., the largest normalized residual test~\cite{Abur_book_2004}) of power system state estimation and thus achieve the stealthiness, if the following conditions hold~\cite{LR_2011}:
\begin{enumerate}
    \item the load measurement injection $\Delta \boldsymbol{D} = \{ \Delta D_d \}$ satisfy the power balance~\eqref{eq_FDI_Ddsum} and the injection limitation~\eqref{eq_FDI_DdLimit};
    \item the line flow measurement injection $\Delta \boldsymbol{F}$ satisfy the power flow equation and limits~\eqref{eq_FDI_FmLimit}.
\end{enumerate}
\begin{alignat}{1}
    &\boldsymbol{1}^T \Delta \boldsymbol{D} = 0 \label{eq_FDI_Ddsum} \\
    &-\boldsymbol{\delta} \circ \boldsymbol{\tau} \circ \boldsymbol{D} \leq \Delta \boldsymbol{D} \leq \boldsymbol{\delta} \circ \boldsymbol{\tau} \circ \boldsymbol{D} \label{eq_FDI_DdLimit} \\
    &-M \boldsymbol{\varepsilon} \leq \Delta \boldsymbol{F} = -\boldsymbol{S} \, \boldsymbol{U_D} \, \Delta \boldsymbol{D} \leq M \boldsymbol{\varepsilon} \label{eq_FDI_FmLimit}
\end{alignat}
where $\circ$ denotes the element-wise product.
$M$ is a sufficiently large constant.
$\tau_d \in \boldsymbol{\tau}$ represents the FDI attacking ability on the $d$-th load measurement~\cite{LR_2011, Xuan2014LR}.

$\boldsymbol{\delta} = \{ \delta_d \}$ and $\boldsymbol{\varepsilon} = \{ \varepsilon_n \}$ are binary variables.
$\delta_d=1$ (or $\varepsilon_n=1$) represents that the $d$-th load (or the $n$-th line flow) is not equipped with measurement protection, indicating that it could be possibly tampered by the FDI cyber-attack.
Note that the cyber-defense resources, e.g., the measurement protection, are usually limited~\cite{ENTSO_2020} and thus should be optimally placed in the system planning stage.

By performing the stealthy FDI cyber-attack~\eqref{eq_FDI_Ddsum}-\eqref{eq_FDI_FmLimit},
the consequent line and load measurement are maliciously and stealthily altered, yielding
\begin{equation} \label{eq_LineLimit_FDI}
    -\overline{\boldsymbol{F}} \leq  \boldsymbol{S} \, (\boldsymbol{U_G} \, \boldsymbol{G} - \boldsymbol{U_D} \, (\boldsymbol{D} + \Delta \boldsymbol{D})) \leq \overline{\boldsymbol{F}}
\end{equation}

As a result, the compromised power system OPF is misled and thus gives inappropriate generation dispatch $\boldsymbol{G}$.
Since this paper focuses on the FDI attack-induced line overloading, the attacking objectives are the maximum overloading of each line (e.g., the $n$-th line) in two opposite directions, yielding the line overloading-oriented FDI cyber-attack~\cite{HminChe_TSG2019}:
\begin{equation} \label{eq_MaxMinObj}
    H_n / V_n = \max / \min \boldsymbol{S}_n \, \boldsymbol{U_D} \, \Delta \boldsymbol{D} \quad \quad \quad \quad \quad \quad
\end{equation}
~~~~~~~~~~~~~~~~~~~~~~~~~~~~over~~$\Delta \boldsymbol{D} \in \mathbb{R}^D$

~~~~~~~~~~~~~~~~~~~~~~~~~~~s.t.~~\eqref{eq_FDI_Ddsum},~\eqref{eq_FDI_DdLimit},~\eqref{eq_FDI_FmLimit}

\noindent
where $\boldsymbol{H} = \{ H_n \}$ and $\boldsymbol{V} = \{ V_n \}$ represent the maximum line overloading in two opposite line flow directions, respectively.
$\boldsymbol{S}_n$ represents the $n$-th row vector of $\boldsymbol{S}$.

\section{A Preventive Cyber-Defense Strategy to Minimize FDI Attack-Induced Region} \label{se_reshape}

\begin{figure}[!b]
    \centering
    \includegraphics[width=8.8cm]{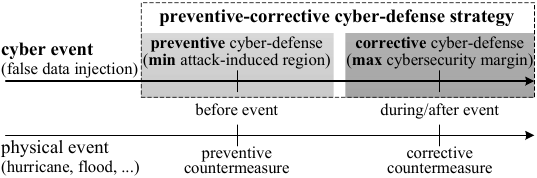}\\
    \vspace{-7pt}
    \caption{Timeline of the preventive and corrective countermeasures before and during/after a cyber or physical event}
    \label{fig_PreCor}
\end{figure}

As shown in Fig.~\ref{fig_PreCor}, this paper proposes a preventive-corrective cyber-defense strategy, which minimizes the FDI attack-induced region in a preventive manner and then maximizes the cybersecurity margin in a corrective manner.
As the first step, this section proposes a preventive cyber-defense strategy to quantify and minimize the volume of the FDI attack-induced region, which is achieved by optimal placement of measurement protection in the system planning stage.

\subsection{Definition and Geometric Illustration of Regions}

Several regions that are defined in the generation injection space, i.e., $\boldsymbol{G}$ space, are given as follows.

\vspace{2pt}
\noindent
\textit{Definition 1. (Security and Insecurity Region)}
The security region $\Theta^{\text{security}}$~\cite{FW1982} and insecurity region $\Theta^{\text{insecurity}}$ of a power system are defined as:
\begin{alignat}{1}
    \Theta^{\text{security}} &:= \{\boldsymbol{G} \mid \eqref{eq_powerbalance},~\eqref{eq_Genlimit},~\eqref{eq_Linelimit} \} \label{eq_defi_security} \\
    \Theta^{\text{insecurity}} &:= \{\boldsymbol{G} \mid \eqref{eq_powerbalance} \} \backslash  \Theta^{\text{security}} \label{eq_defi_insecurity}
\end{alignat}

Note that
the stealthy FDI cyber-attack is able to cause any magnitude of line overloading within the damaging boundary $\boldsymbol{V}$ and $\boldsymbol{H}$.
Thus, the lower and upper bounds of the line flow $\boldsymbol{F}$, i.e., $-\overline{\boldsymbol{F}}$ and $\overline{\boldsymbol{F}}$, are maliciously and stealthily expanded by $\boldsymbol{V}$ and $\boldsymbol{H}$, respectively.
This yields
\begin{equation} \label{eq_CRcons}
    \boldsymbol{V} -\overline{\boldsymbol{F}} \leq \boldsymbol{F} = \boldsymbol{S} \, (\boldsymbol{U_G} \, \boldsymbol{G} - \boldsymbol{U_D} \, \boldsymbol{D})  \leq \boldsymbol{H} + \overline{\boldsymbol{F}}
\end{equation}

As a result, the line flow constraints~\eqref{eq_Linelimit} are violated if the consequent line flow $\boldsymbol{F}$ fall in the following range:
\begin{equation} \label{eq_CIRcons}
    \boldsymbol{V} -\overline{\boldsymbol{F}} \leq \boldsymbol{F} \leq -\overline{\boldsymbol{F}} \quad \text{or} \quad \overline{\boldsymbol{F}} \leq \boldsymbol{F} \leq \boldsymbol{H} + \overline{\boldsymbol{F}}
\end{equation}

Then we have the following two regions related to the stealthy FDI cyber-attacks.

\vspace{2pt}
\noindent
\textit{Definition 2. (\textbf{FDI Attack-Induced Region})}
Consider a power system subjected to~\eqref{eq_powerbalance}-\eqref{eq_Linelimit} and under stealthy FDI cyber-attack~\eqref{eq_FDI_Ddsum}-\eqref{eq_FDI_FmLimit}, the FDI attack-reachable region (ARR) $\Omega^{\text{ARR}}$ and FDI attack-induced region (AIR) $\Omega^{\text{AIR}}$ are defined as:
\begin{alignat}{1}
    \Omega^{\text{ARR}} &:= \{\boldsymbol{G} \mid \eqref{eq_powerbalance},~\eqref{eq_Genlimit},~\eqref{eq_CRcons} \} \label{eq_defi_CRR} \\
    \Omega^{\text{AIR}} &:= \{\boldsymbol{G} \mid \eqref{eq_powerbalance},~\eqref{eq_Genlimit},~\eqref{eq_CIRcons} \} \subset \Omega^{\text{ARR}} \label{eq_defi_CIR}
\end{alignat}

\begin{figure}[!b]
    \centering
    \includegraphics[width=8.8cm]{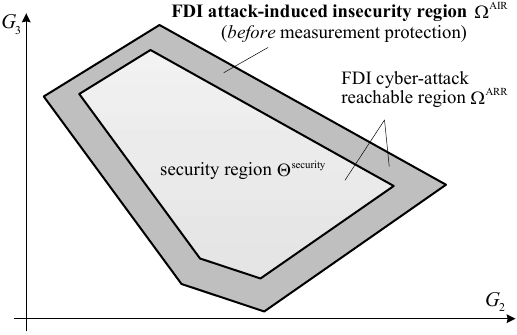}\\
    \vspace{-7pt}
    \caption{Conceptual illustration of the security region $\Theta^{\text{security}}$, the FDI attack-reachable region (ARR) $\Omega^{\text{ARR}}$, and the FDI attack-induced region (AIR) $\Omega^{\text{AIR}}$ of a three-generator power system in $\boldsymbol{G}$ space}
    \label{fig_regionRelation}
\end{figure}

According to the definitions, $\Omega^{\text{ARR}}$ in~\eqref{eq_defi_security} and $\Theta^{\text{security}}$ in~\eqref{eq_defi_CRR} are both polytopes in $\boldsymbol{G}$ space.
Thus, the aforementioned regions of a three-generator power system (with two independent generators) are conceptually illustrated in Fig.~\ref{fig_regionRelation}.

\subsection{FDI Unattackable Lines}

To start with, we have the following proposition for the maximum FDI attack-induced line overloading.

\vspace{2pt}
\noindent
\textit{Proposition~1. (maximum FDI Attack-Induced Line Overloading)} 
$\boldsymbol{H}$ and $\boldsymbol{V}$ solved by~\eqref{eq_MaxMinObj} satisfy the following two properties:
\begin{alignat}{1}
    &\boldsymbol{H} = -\boldsymbol{V} \label{eq_P1_1} \\
    &\boldsymbol{H} \ge 0, \ \boldsymbol{V} \leq 0 \label{eq_P1_2}
\end{alignat}

\vspace{2pt}
\noindent
\textit{Proof.}
$H_n = \max \boldsymbol{S}_n  \boldsymbol{U_D}  \Delta \boldsymbol{D} = \min - \boldsymbol{S}_n  \boldsymbol{U_D}  \Delta \boldsymbol{D} = -V_n$, yielding~\eqref{eq_P1_1}.
Then, $H_n \ge V_n$ and~\eqref{eq_P1_1} yield~\eqref{eq_P1_2}.~~~~~~~~~~~~$\square$

If the equality in~\eqref{eq_P1_2} holds, we have the FDI unattackable line that is defined as follows.

\vspace{2pt}
\noindent
\textit{Definition 3. (FDI Unattackable Line)}
The $n$-th line is termed as FDI unattackable if there exists no non-zero stealthy FDI cyber-attack~\eqref{eq_MaxMinObj} on the $n$-th line.
That is, for any $\Delta \boldsymbol{D}$ that satisfies~\eqref{eq_FDI_Ddsum} and~\eqref{eq_FDI_DdLimit}, we have
\begin{equation} \label{eq_Hn0Ln0}
    H_n = V_n = 0
\end{equation}

\vspace{3pt}
\noindent
\textit{Remark~1.}
The previous studies~\cite{bobba2010detecting, TII2021_GraphSensor, TSG2014_GraphSensor, TSG2013_Game_Sensor, TII2015_Greedy_Sensor, TSG2019_Comb_Sensor} focus on protecting a selected measurement subset to globally and completely prevent any stealthy FDI cyber-attack, indicating that~\eqref{eq_Hn0Ln0} holds for all $n$.
This relies on a strong assumption that there exist sufficient cyber-defense resources for measurement protection.
By comparison, this paper investigates how to optimally place limited cyber-defense resources that may be insufficient to prevent all the stealthy FDI cyber-attacks, leading to 1) reduced but non-zero $\boldsymbol{H}$ and $\boldsymbol{L}$; 2) or $H_n = V_n = 0$ for a subset of lines.

Let $d_n'$ and $d_n''$ denote the two terminal buses of the $n$-th line.
Let $\mathcal{N}_d, d \in \{d_n', d_n''\}$, denotes the set of all adjacent lines of the $n$-th line via the terminal bus $d$.

\vspace{3pt}
\noindent
\textit{Proposition~2. (\textbf{A Sufficient Condition for Constructing FDI Unattackable Lines with Measurement Protection})}
Given a measurement protection strategy $\boldsymbol{\delta}$ and $\boldsymbol{\varepsilon}$,
the $n$-th line is unattackable if there exists $d \in \{d', d'' \}$ such that
\begin{equation} \label{eq_unattack_d}
    \delta_d = 0
\end{equation}
and
\begin{equation} \label{eq_unattack_l}
    \sum_{l \in \mathcal{N}_d} \varepsilon_l = 0 \ \text{or} \ \mathcal{N}_d = \emptyset
\end{equation}
hold.

\vspace{1pt}
\noindent
\textit{Proof.}
See Appendix~\ref{ap_unattack}.

\subsection{Quantifying the Volume of the FDI Attack-Induced Region}

Proposition~1 indicates that $\Theta^{\text{security}}$ is a subset of $\Omega^{\text{ARR}}$, yielding $\Omega^{\text{ARR}} \supseteq \Theta^{\text{security}}$.
From a geometric viewpoint, $\Theta^{\text{security}}$ is \textit{similar} to $\Omega^{\text{ARR}}$ in $\boldsymbol{G}$ space in Fig.~\ref{fig_regionRelation}.
Thus, the FDI attack-induced region $\Omega^{\text{AIR}}$ is the difference of the two polytopes, yielding
\begin{equation}
    \Omega^{\text{AIR}} = \Omega^{\text{ARR}} \backslash \Theta^{\text{security}} = \Omega^{\text{ARR}} \cap \Theta^{\text{insecurity}} 
\end{equation}

Specifically, if all lines are unattackable, i.e., $\boldsymbol{H} = \boldsymbol{V} = \boldsymbol{0}$, then $\Omega^{\text{ARR}}$ would shrink to $\Theta^{\text{security}}$ and thus $\Omega^{\text{AIR}}$ would be an empty set.

Power lines have different $\overline{F_n}$, implying different levels of tolerance towards overloading.
That is, for the same $H_n$ solved by~\eqref{eq_MaxMinObj}, a line with smaller $\overline{F_n}$ would be more dangerous than a line with larger $\overline{F_n}$.
In this regard, the weighted sum of $H_n$ with weights $1/\overline{F_n}$ quantifies the overall normalized attack-induced damages: 
\begin{equation} \label{eq_benefitCal}
    \sum_{n \in \mathcal{L}} H_n / \overline{F_n}
\end{equation}
where $\mathcal{L}$ denotes the set of all lines.
From a geometric viewpoint, $H_n$ indicates how large $\Theta^{\text{security}}$ is maliciously expanded to $\Omega^{\text{ARR}}$ with respect to the $n$-th line.
In this regard,~\eqref{eq_benefitCal} equivalently quantifies the volume of FDI attack-induced region, i.e., $\Omega^{\text{AIR}}$ in Fig.~\ref{fig_regionRelation}.

\subsection{Two-Objective Optimization for Shaping the FDI Attack-Induced Region}

\begin{figure}[!b]
    \centering
    \includegraphics[width=8.8cm]{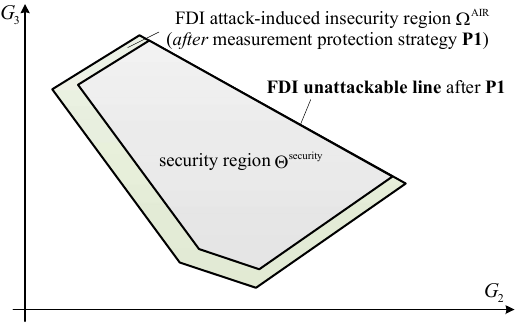}\\
    \vspace{-7pt}
    \caption{Consequent FDI attack-induced region $\Omega^{\text{AIR}}$ after implementing the preventive measurement protection strategy P1 on a three-generator power system}
    \label{fig_reducedCIR}
\end{figure}

Given the system topology and parameters,
the volume of the FDI attack-induced region is determined by two factors:
\begin{itemize}
    \item the FDI cyber-attack abilities $\boldsymbol{\tau}$~\cite{LR_2011, Xuan2014LR} (i.e., the limitation of cyber-attack costs), whose increment increases $\boldsymbol{H}$ and thus increases the region volume. We leave it for future investigation since this paper mainly focuses on the costs and benefits of cyber-defense strategies.
    \item the measurement protection strategies (i.e., $\boldsymbol{\delta}$ and $\boldsymbol{\varepsilon}$), whose increment and optimal placement could decrease $\boldsymbol{H}$ and thus decrease the region volume.
    This is conceptually illustrated in Fig.~\ref{fig_reducedCIR}.
\end{itemize}

Since a smaller FDI attack-induced region volume is beneficial for the power system,
the cyber-defense benefit of the proposed preventive cyber-defense strategy is defined as
\begin{equation} \label{eq_BeneReshape}
    B^{\text{AIR}} = - \sum_{n \in \mathcal{L}} H_n / \overline{F_n}
\end{equation}

The cyber-defense cost can be quantified by the sum of all the measurement protection in load and line flow:
\begin{equation} \label{eq_costCal}
    C^{\text{AIR}} = \sum_{d \in \mathcal{D}} (1 - \delta_d) + \sum_{n \in \mathcal{L}} (1 - \varepsilon_n) \leq \overline{C^{\text{AIR}}}
\end{equation}
where $\overline{C^{\text{AIR}}}$ indicates the limited cyber-defense resource that may be insufficient to completely prevent all the stealthy FDI cyber-attacks.

The cyber defenders in system planning aim to optimize two objectives.
One objective is to maximize the cyber-defense benefit $B^{\text{AIR}}$, i.e., minimizing the maximum attack-induced line overloading of each line.
The other objective is to minimize the cyber-defense cost $C^{\text{AIR}}$, i.e., minimizing the number of measurement protections.
This yields a bi-objective bi-level (one-leader-multi-follower) optimization problem (denoted as P1$'$).
\begin{alignat}{1}
    \text{P1}': \ &\min \ \{ - B^{\text{AIR}},~C^{\text{AIR}} \} \notag \\
    &\text{over}  \quad \boldsymbol{\delta} \in \{0, 1\}^D, \boldsymbol{\varepsilon} \in \{0, 1\}^{L} \notag \\
    &\text{s.t.} \quad \eqref{eq_BeneReshape},~\eqref{eq_costCal} \notag \\
    &\quad \quad H_n = \max \ \boldsymbol{S}_n \, \boldsymbol{U_D} \, \Delta \boldsymbol{D}^{n} \quad \forall n \in \mathcal{L} \label{eq_HnmaxLow} \\
    &\quad \quad \text{over} \ \ \Delta \boldsymbol{D}^{n} \in \mathbb{R}^D \quad \forall n \in \mathcal{L} \label{eq_HnmaxLowOver} \\
    &\quad \quad \quad \text{s.t.}~~\boldsymbol{1}^T \Delta \boldsymbol{D}^{n} = 0 \label{eq_FDI_Ddsum_Dn} \\
    &\quad \quad \quad \quad -\boldsymbol{\delta} \circ \boldsymbol{\tau} \circ \boldsymbol{D} \leq \Delta \boldsymbol{D}^{n} \leq \boldsymbol{\delta} \circ \boldsymbol{\tau} \circ \boldsymbol{D} \label{eq_FDI_DdLimit_Dn} \\
    &\quad \quad \quad \quad -M \boldsymbol{\varepsilon} \leq \boldsymbol{S} \, \boldsymbol{U_D} \, \Delta \boldsymbol{D}^{n} \leq M \boldsymbol{\varepsilon} \label{eq_FDI_FmLimit_Dn}
\end{alignat}
where
$\Delta \boldsymbol{D}^{n}$ denotes the FDI cyber-attack injections with respect to the $n$-th line.
Each follower problem indicates an optimal FDI cyber-attack strategy for each line with a certain measurement protection strategy, while the leader problem aims to optimize two objectives.
The two objectives conflict with each other, since using less measurement protection leads to a larger FDI attack-induced region volume.
Thus, there does not exist an optimal point such that the two conflicting objectives are concurrently optimized. Instead, we can achieve the Pareto optimal defined as follows.

\vspace{2pt}
\noindent
\textit{Definition~4. (Pareto Optimal)}~\cite{MulObj_SIAM1998}
For a general multi-objective optimization problem $\min_{x \in X} \{ f_1(x), ..., f_j(x), ... \}$,
a point $x^* \in X$ is Pareto optimal if and only if there does not exist another point $x \in X$ such that $f_j(x) \le f_j(x^*) \ \forall j$ and $f_j(x) < f_j(x^*)$ for at least one $j$.

The weighted sum method~\cite{MulObj_weighted_2010} is adopted to solve the bi-objective bi-level problem P1$'$ and obtain the corresponding Pareto optimal front. Therein, the two objectives are scalarized into a single objective by adding a user-supplied weight $\omega^{\text{AIR}}$ in an objective, yielding a single-objective bi-level problem P1:
\begin{alignat}{1}
    &\text{P1}: \ \min_{\boldsymbol{\delta} \in \{0, 1\}^D, \boldsymbol{\varepsilon} \in \{0, 1\}^{L}} \ - B^{\text{AIR}} + \omega^{\text{AIR}} C^{\text{AIR}} \notag \\
    &\text{s.t.} \quad \eqref{eq_BeneReshape}-\eqref{eq_FDI_FmLimit_Dn} \notag
\end{alignat}
where $\omega^{\text{AIR}}$ represents the weight coefficient to balance the cyber-defense cost and benefit.
The Pareto optimal front of P1$'$ can be obtained by varying $\omega^{\text{AIR}}$.
As a result, the Pareto optimal front enables the system operators with a preventive cyber-defense strategy, which indicates how to maximize $B^{\text{AIR}}$ with respect to any given measurement protection budget.

The proposed single-objective bi-level problem P1 will be converted into a single-objective single-level problem P2 in Section~\ref{se_reshapeSolution}.
After solving the proposed preventive cyber-defense strategy P1/P2,
the consequent maximum line-overloading will be delivered to the subsequent corrective cyber-defense strategy P3 in Section~\ref{se_redispatch}.

\subsection{Model Extension with Load Variations}

The proposed preventive cyber-defense strategy P1 in system planning can be extended with consideration of load variations.
Assume the loads $\boldsymbol{D}$ follow the distribution of $\xi$.
P1 remains a bi-objective bi-level problem in a stochastic manner:
\begin{alignat}{1}
    &\min_{\boldsymbol{\delta} \in \{0, 1\}^D, \boldsymbol{\varepsilon} \in \{0, 1\}^{L}} \ - B^{\text{AIR}}(\xi) + \omega^{\text{AIR}} C^{\text{AIR}} \notag \\
    &\quad \quad \quad \text{s.t.} \quad \eqref{eq_costCal}  \notag \\
    &\quad \quad \quad B^{\text{AIR}}(\xi) = - \sum_{n \in \mathcal{L}} \mathbb{E}[H_n(\xi)] / \overline{F_n} \label{eq_BAIR_xi} \\
    &\quad \quad \quad H_n(\xi) = \max_{\Delta \boldsymbol{D}^{n} \in \mathbb{R}^D} \ \boldsymbol{S}_n \, \boldsymbol{U_D} \, \Delta \boldsymbol{D}^{n} \quad \forall n \in \mathcal{L} \label{eq_HnmaxLow_xi} \\
    &\quad \quad \quad  \text{s.t.} \quad \eqref{eq_FDI_Ddsum_Dn},~\eqref{eq_FDI_FmLimit_Dn} \notag \\
    &\quad \quad \quad  \quad \quad -\boldsymbol{\delta} \circ \boldsymbol{\tau} \circ \boldsymbol{D}(\xi) \leq \Delta \boldsymbol{D}^{n} \leq \boldsymbol{\delta} \circ \boldsymbol{\tau} \circ \boldsymbol{D}(\xi) \label{eq_deltaD_xi}
\end{alignat}
where $\mathbb{E}[H_n(\xi)]$ represents the expected value of the maximum attack-induced line overloading with respect to $\boldsymbol{D}(\xi)$.
Since the stochastic properties are beyond the main scope of this paper, we leave it for future investigation.

\section{Solution to the Preventive Cyber-Defense Strategy} \label{se_reshapeSolution}

To solve the proposed cost-effective measurement protection strategy, which is a bi-level problem in P1, this section reformulates P1 into a single-level MILP problem.
Moreover, the closed-form lower bounds of the big-M are addressed to help efficiently solve the MILP problem.

\subsection{Problem Reformulation}

By applying the Karush-Kuhn-Tucker (KKT) optimality condition and the Fortuny-Amat mixed-integer reformulation~\cite{fortuny1981}, the original one-leader-multi-follower problem P1 is converted into a single-level mixed-integer linear programming (MILP) problem (denoted as P2).
\begin{alignat}{1}
    \text{P2}: \ &\min -\ B^{\text{AIR}} + \omega^{\text{AIR}} C^{\text{AIR}} \label{eq_objAIR_P2} \\
    &\text{over}  \quad \boldsymbol{\delta} \in \{0, 1\}^D, \boldsymbol{\varepsilon}  \in \{0, 1\}^{L} \notag \\
    &\quad \quad \ \Delta \boldsymbol{D}^{n}  \in \mathbb{R}^D    \quad \forall n \in \mathcal{L}    \notag
    \end{alignat}
    \begin{alignat}{1}
    &\text{s.t.} \quad   \eqref{eq_BeneReshape},~\eqref{eq_costCal} \notag \\
    &\quad \quad \, \eqref{eq_FDI_Ddsum_Dn},~\eqref{eq_FDI_DdLimit_Dn},~\eqref{eq_FDI_FmLimit_Dn} \quad  \forall n \in \mathcal{L} \notag
\end{alignat}
\begin{equation}
    H_n = \boldsymbol{S}_n \, \boldsymbol{U_D} \, \Delta \boldsymbol{D}^{n} \quad \forall n \in \mathcal{L}
\end{equation}

\vspace{-10pt}

\begin{subequations}
    \begin{align}
        &\boldsymbol{0}^T = \boldsymbol{S}_n \, \boldsymbol{U_D} + \lambda^{n} \boldsymbol{1}^T + (\overline{\boldsymbol{\alpha}^{n}} - \underline{\boldsymbol{\alpha}^{n}})^T \notag \\
        &\quad \quad +  (\overline{\boldsymbol{\beta}^{n}} - \underline{\boldsymbol{\beta}^{n}})^T \, \boldsymbol{S} \, \boldsymbol{U_D} \\
        &\lambda^n \in \mathbb{R} \\
        &\forall n \in \mathcal{L} \notag
    \end{align}
\end{subequations}

\vspace{-10pt}

\begin{subequations}  \label{eq_alphagroup}
    \begin{align} 
    &\underline{\boldsymbol{\alpha}^n} \leq K \boldsymbol{u}_{\underline{\boldsymbol{\alpha}^n}}, \ \boldsymbol{\delta} \circ \boldsymbol{\tau} \circ \boldsymbol{D} + \Delta \boldsymbol{D}^n \leq K (\boldsymbol{1} - \boldsymbol{u}_{\underline{\boldsymbol{\alpha}^n}}) \\
    &\overline{\boldsymbol{\alpha}^n} \leq K \boldsymbol{u}_{\overline{\boldsymbol{\alpha}^n}}, \ \boldsymbol{\delta} \circ \boldsymbol{\tau} \circ \boldsymbol{D} - \Delta \boldsymbol{D}^n \leq K (\boldsymbol{1} - \boldsymbol{u}_{\overline{\boldsymbol{\alpha}^n}}) \\
    &\underline{\boldsymbol{\alpha}^n}, \overline{\boldsymbol{\alpha}^n} \ge \boldsymbol{0}, \, \underline{\boldsymbol{\alpha}^n}, \overline{\boldsymbol{\alpha}^n}  \in \mathbb{R}^D, \, \boldsymbol{u}_{\underline{\boldsymbol{\alpha}^n}}, \boldsymbol{u}_{\overline{\boldsymbol{\alpha}^n}} \in \{0,1 \}^D \\
    &\boldsymbol{u}_{\underline{\boldsymbol{\alpha}^n}} + \boldsymbol{u}_{\overline{\boldsymbol{\alpha}^n}} \leq \varepsilon_n \boldsymbol{1} \label{eq_alphagroup_uu}  \\
    &\forall n \in \mathcal{L} \notag
    \end{align}
\end{subequations}

\vspace{-10pt}

\begin{subequations} \label{eq_betagroup}
    \begin{align}
    &\underline{\boldsymbol{\beta}^n} \leq N \boldsymbol{u}_{\underline{\boldsymbol{\beta}^n}}, \ M \boldsymbol{\varepsilon} + \boldsymbol{S} \, \boldsymbol{U_D} \, \Delta \boldsymbol{D}^{n} \leq N (\boldsymbol{1} - \boldsymbol{u}_{\underline{\boldsymbol{\beta}^n}}) \\
    &\overline{\boldsymbol{\beta}^n} \leq N \boldsymbol{u}_{\overline{\boldsymbol{\beta}^n}}, \ M \boldsymbol{\varepsilon} - \boldsymbol{S} \, \boldsymbol{U_D} \, \Delta \boldsymbol{D}^{n} \leq N (\boldsymbol{1} - \boldsymbol{u}_{\overline{\boldsymbol{\beta}^n}}) \\
    &\underline{\boldsymbol{\beta}^n}, \overline{\boldsymbol{\beta}^n} \ge \boldsymbol{0}, \, \underline{\boldsymbol{\beta}^n}, \overline{\boldsymbol{\beta}^n}  \in \mathbb{R}^{L}, \, \boldsymbol{u}_{\underline{\boldsymbol{\beta}^n}}, \boldsymbol{u}_{\overline{\boldsymbol{\beta}^n}} \in \{0,1 \}^{L} \\
    &\boldsymbol{u}_{\underline{\boldsymbol{\beta}^n}} + \boldsymbol{u}_{\overline{\boldsymbol{\beta}^n}} + \boldsymbol{\varepsilon} = \boldsymbol{1} \label{eq_betagroup_equality} \\
    &\boldsymbol{u}_{\underline{\boldsymbol{\beta}^n}} \le \boldsymbol{u}_{\overline{\boldsymbol{\beta}^n}} \label{eq_betagroup_uu}  \\
    &\forall n \in \mathcal{L} \notag
    \end{align}
\end{subequations}

\vspace{-10pt}

\begin{subequations}
    \begin{align}
    &0 < M < N \\
    &0 < K
    \end{align}
\end{subequations}
where
$\lambda^n$ is the Lagrange multiplier associated with~\eqref{eq_FDI_Ddsum_Dn}.
$\underline{\boldsymbol{\alpha}^n} = \{ \underline{\alpha^n_d} \}$ and $\overline{\boldsymbol{\alpha}^n} = \{ \overline{\alpha^n_d} \}$ are the Lagrange multiplier vectors associated with~\eqref{eq_FDI_DdLimit_Dn},
whose complementary slackness conditions are represented by introducing the binary variable vectors $\boldsymbol{u}_{\underline{\boldsymbol{\alpha}^n}} = \{ u_{\underline{\alpha^n_d}} \}$ and $\boldsymbol{u}_{\overline{\boldsymbol{\alpha}^n}} = \{ u_{\overline{\alpha^n_d}} \}$, respectively.
$\underline{\boldsymbol{\beta}^n} = \{ \underline{\beta^n_l} \}$ and $\overline{\boldsymbol{\beta}^n} = \{ \overline{\beta^n_l} \}$ are the Lagrange multiplier vectors associated with~\eqref{eq_FDI_FmLimit_Dn},
whose complementary slackness conditions are represented by introducing the binary variable vectors $\boldsymbol{u}_{\underline{\boldsymbol{\beta}^n}} = \{ u_{\underline{\beta^n_l}} \}$ and $\boldsymbol{u}_{\overline{\boldsymbol{\beta}^n}} = \{ u_{\overline{\beta^n_l}} \}$, respectively.
$M$, $N$, and $K$ are sufficiently large constants.

Note that~\eqref{eq_alphagroup_uu} and~\eqref{eq_betagroup_uu} help avoid unnecessary enumeration in the solution space and thus improve the computational performance.
Note that~\eqref{eq_betagroup_equality} ensures that~\eqref{eq_FDI_FmLimit_Dn} holds either with or without line measurement protection.

\vspace{3pt}
\noindent
\textit{Remark~2.}
Proposition~2 implies that protecting line measurements may not be the best decision for P2 to reduce the maximum line overloading.
Instead, to reduce the sum of the maximum line overloading of all lines, i.e., $\sum_n H_n / \overline{F_n}$ in~\eqref{eq_BeneReshape},
it could be more efficient to place the measurement protection on loads rather than on lines.
This is because a protected line measurement in P2 only reduces the maximum overloading of the line itself, while a protected load measurement in P2 could reduce the maximum overloading of multiple lines that are interconnected to the load bus.

\subsection{Lower Bounds of the Big-M}

There are two kinds of large constants in P2:
\begin{itemize}
    \item $M$ is introduced to set a large range for the FDI line flow injections $\Delta \boldsymbol{F}$ in~\eqref{eq_FDI_FmLimit_Dn} if there exists no measurement protection.
    \item $K$ and $N$, which are also known as the big-M~\cite{fortuny1981}, are introduced to avoid violating the original constraints~\eqref{eq_FDI_DdLimit_Dn} and~\eqref{eq_FDI_FmLimit_Dn} when the corresponding Lagrange multipliers are not zero~\cite{BigM_TPS2019}.
\end{itemize}

Inappropriate values of the big-M would deteriorate the convergence and optimum of the proposed MILP problem P2~\cite{BigM_TPS2019}.
Specifically, too large values of $M$, $N$, and $K$ would make the MILP problem P2 computationally intractable and hard to converge.
By contrast, too small values of $M$, $N$, and $K$ would violate the original constraints~\eqref{eq_FDI_DdLimit_Dn} and~\eqref{eq_FDI_FmLimit_Dn}, leading to failures of convergence.
Thus, closed-form lower bounds for the sufficiently large constants ($M$, $N$, and $K$) are addressed as follows.

\vspace{3pt}
\noindent
\textit{Proposition~3. (\textbf{Closed-Form Lower Bounds of the Big-M})}
$M$, $N$, and $K$ in P2 are sufficiently large (i.e., \eqref{eq_FDI_DdLimit_Dn} and~\eqref{eq_FDI_FmLimit_Dn} are not violated) if
\begin{alignat}{1}
    M &\ge \| \ | \boldsymbol{S} \boldsymbol{U_D} | \ ( \boldsymbol{\tau} \circ \boldsymbol{D} ) \  \|_{\infty} \label{eq_bigM_M} \\
    N &\ge M + \| \ | \boldsymbol{S} \boldsymbol{U_D} | \ ( \boldsymbol{\tau} \circ \boldsymbol{D} ) \  \|_{\infty} \label{eq_bigM_N} \\
    K &\ge 2 \| \ \boldsymbol{\tau} \circ \boldsymbol{D} \  \|_{\infty} \label{eq_bigM_K}
\end{alignat}

\vspace{1pt}
\noindent
\textit{Proof.}
See Appendix~\ref{ap_bigM}.

\subsection{Effectiveness and Cost-Effectiveness of the Proposed Preventive Cyber-Defense Strategy P2}

After solving the proposed preventive cyber-defense strategy P2, we can obtain:
\begin{itemize}
    \item the smallest number and allocation of measurement protection that can completely eliminate the FDI attack-induced region, which is widely addressed by previous studies~\cite{bobba2010detecting, TII2021_GraphSensor, TSG2014_GraphSensor, TSG2013_Game_Sensor, TII2015_Greedy_Sensor, TSG2019_Comb_Sensor}.
    \item the optimal placement of insufficient measurement protection to minimize the volume of the FDI attack-induced region.
    \item the cyber-defense benefits with respect to any given cyber-defense cost budget $\overline{C^{\text{AIR}}}$.
    \item the \textit{cyber-defense marginal benefit} with respect to the increment of the cyber-defense cost.
\end{itemize}

\section{A Corrective Cyber-Defense Strategy to Maximize Cybersecurity Margin} \label{se_redispatch}

With limited cyber-defense resources,
the proposed preventive cyber-defense strategy P2 may not completely eliminate the FDI attack-induced region $\Omega^{\text{AIR}}$.
Since the power system generation dispatch could be either compromised by cyber-attacks~\cite{LR_2011} or actively adopted as a resilience enhancement tool~\cite{TSG2022_LAA}, this section proposes a corrective cyber-defense strategy to maximize the cybersecurity margin via corrective and cost-effective re-dispatch of operation points.

\subsection{Quantifying Cybersecurity Margin of Operation Points}

After implementing the proposed preventive cyber-defense strategy P2 in system planning, the consequent maximum line overloading (denoted as $\boldsymbol{H}^*$) may not equal $\boldsymbol{0}$ due to a limited cyber-defense cost budget.
In this regard, the power system may still face cyber threats during cyber-attack events.
Hence, a corrective countermeasure should be proposed to dispatch the original operation point to a new one with a larger cybersecurity margin, which provides a buffer to avoid attack-induced damages.
To timely implement the corrective cyber-defense strategy, many well-developed cyber-attack detection techniques can be adopted, such as model-based detection methods~\cite{TII2021_GraphSensor, TSG2014_GraphSensor, TSG2013_Game_Sensor, TII2015_Greedy_Sensor, TSG2019_Comb_Sensor} or data-driven detection methods~\cite{FDIDet_Survery_TSG2020}.

\begin{figure}[!b]
    \centering
    \includegraphics[width=8.8cm]{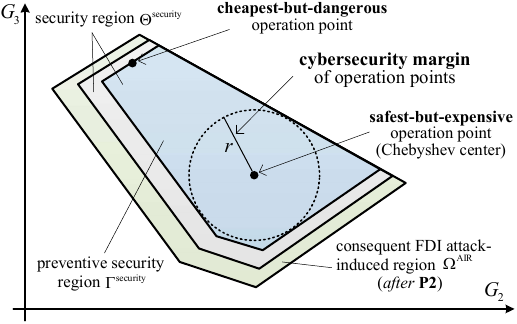}\\
    \vspace{-7pt}
    \caption{Proposed corrective cyber-defense strategy P3 to achieve the trade-off between the safest-but-expensive operation point (i.e., the Chebyshev center) and the cheapest-but-dangerous operation point within the preventive security region}
    \label{fig_Redis}
\end{figure}

Specifically, given the consequent FDI attack-induced region $\Omega^{\text{AIR}}$ (green area in Fig.~\ref{fig_Redis}) that may not be an empty set, power lines still face different levels of cyber threats.
In this regard, the original security region should be shrunk into a preventive security region, which is defined in Definition~6 and represented by the blue area in Fig.~\ref{fig_Redis}.
That is, the preventive security region $\Gamma^{\text{security}}$ is smaller than or equals to the security region $\Theta^{\text{security}}$ as $\boldsymbol{H}^* \ge 0$.
Therein, the cybersecurity margin of an operation point is defined in Definition~7.

\vspace{2pt}
\noindent
\textit{Definition~5. (Preventive Line Limits)}
Given the consequent maximum line overloading $\boldsymbol{H}^*$ after implementing the proposed preventive cyber-defense strategy P2,
the preventive line limits are defined as
\begin{equation} \label{eq_LineLimit_SR}
    -\overline{\boldsymbol{F}} + \boldsymbol{H}^* \leq \boldsymbol{F} = \boldsymbol{S} \, (\boldsymbol{U_G} \, \boldsymbol{G} - \boldsymbol{U_D} \, \boldsymbol{D}) \leq \overline{\boldsymbol{F}} - \boldsymbol{H}^*
\end{equation}
yielding a set of linear inequalities
\begin{equation} \label{eq_AGb}
    \boldsymbol{A} \, \boldsymbol{G} \leq \boldsymbol{b}
\end{equation}
where
\begin{equation} \label{eq_Ab}
    \boldsymbol{A} = \begin{bmatrix}
    \boldsymbol{S} \, \boldsymbol{U_G}  \\
    -\boldsymbol{S} \, \boldsymbol{U_G}
    \end{bmatrix}, \quad
    \boldsymbol{b} = \begin{bmatrix}
    \overline{\boldsymbol{F}} - \boldsymbol{H}^* + \boldsymbol{S} \, \boldsymbol{U_D} \, \boldsymbol{D}  \\
    \overline{\boldsymbol{F}} - \boldsymbol{H}^* - \boldsymbol{S} \, \boldsymbol{U_D} \, \boldsymbol{D}
\end{bmatrix}
\end{equation}

\vspace{2pt}
\noindent
\textit{Definition~6. (\textbf{Preventive Security Region})}
Given the consequent $\boldsymbol{H}^*$ after implementing P2,
the preventive security region is defined as
\begin{equation}
    \Gamma^{\text{security}} := \{\boldsymbol{G} \mid \eqref{eq_powerbalance},~\eqref{eq_Genlimit},~\eqref{eq_LineLimit_SR} \}
\end{equation}

The distance of a given operation point $\boldsymbol{G}$ to the boundaries~\eqref{eq_AGb} is
\begin{equation} \label{eq_AGbri_e}
    q_i = \frac{ | b_i - \boldsymbol{A}_i \boldsymbol{G} | }{\| \boldsymbol{A}_i \|_2}   \quad \forall i
\end{equation}
where $\boldsymbol{A}_i$ denotes the $i$-th row of the matrix $\boldsymbol{A}$.
$b_i$ denotes the $i$-th element of the vector $\boldsymbol{b}$.

\vspace{2pt}
\noindent
\textit{Definition~7. (\textbf{Cybersecurity Margin of an Operation Point})}
The cybersecurity margin of a given operation point $\boldsymbol{G} \in \Gamma^{\text{security}}$, denoted as $r(\boldsymbol{G})$ in Fig.~\ref{fig_Redis}, is defined as the
minimum Euclidean distance to the boundaries of $\Gamma^{\text{security}}$ in~\eqref{eq_AGb}.

\begin{table*}[!t]
    \centering
    \vspace{-6pt}
    \caption{The Trade-Off between Cyber-Defense Cost and Benefit of the Proposed Preventive-Corrective Cyber-Defense Strategy}
    \label{table_cyberCostBene}
\begin{tabular}{ccc}
\hline \hline
& preventive cyber-defense strategy P2  & corrective cyber-defense strategy P3  \\ \hline \hline
\begin{tabular}[c]{@{}c@{}}system\\ stage\end{tabular} & system planning & system operation \\ \hline
\begin{tabular}[c]{@{}c@{}}cyber-defense\\ cost\end{tabular}              & \begin{tabular}[c]{@{}c@{}}number of measurement protections\\ $C^{\text{AIR}} = \sum_{d \in \mathcal{D}} (1 - \delta_d) + \sum_{n \in \mathcal{L}} (1 - \varepsilon_n)$\end{tabular} & \begin{tabular}[c]{@{}c@{}}additional defense-induced operation cost\\ $C^G = \boldsymbol{c}^T \, \boldsymbol{G}$\end{tabular}                                 \\ \hline
\begin{tabular}[c]{@{}c@{}}cyber-defense\\ benefit\end{tabular}           & \begin{tabular}[c]{@{}c@{}}opposite value of the FDI attack-induced region volume\\ $B^{\text{AIR}} = -\sum_{n \in \mathcal{L}} H_n / \overline{F_n}$ \\ (a measure of the set $\Omega^{\text{AIR}}$)\end{tabular}   & \begin{tabular}[c]{@{}c@{}}cybersecurity margin\\ $B^G = r(\boldsymbol{G})$ \\ (a measure of the element $\boldsymbol{G}$ in the set $\Gamma^{\text{security}}$) \end{tabular}           \\ \hline
\begin{tabular}[c]{@{}c@{}}cyber-defense\\ decision variable\end{tabular} & \begin{tabular}[c]{@{}c@{}}optimal measurement protection on loads and lines\\ $\boldsymbol{\delta} \in \{0, 1\}^D, \boldsymbol{\varepsilon}  \in \{0, 1\}^{L}$\end{tabular}        & \begin{tabular}[c]{@{}c@{}}optimal generation dispatch\\ $\boldsymbol{G} \in \mathbb{R}^G$\end{tabular}                             \\ \hline
\begin{tabular}[c]{@{}c@{}} cyber-defense\\ bi-objective \end{tabular}         & \begin{tabular}[c]{@{}c@{}}max cyber-defense benefits and min cyber-defense costs \\ $\min \{ -B^{\text{AIR}},~C^{\text{AIR}} \} $ \end{tabular}                                            & \begin{tabular}[c]{@{}c@{}}max cyber-defense benefits and min cyber-defense costs\\ $\min \{ -B^G,~C^G \}$ \end{tabular} \\ \hline
\begin{tabular}[c]{@{}c@{}}information\\ transmission\end{tabular} & \multicolumn{2}{c}{deliver the consequent maximum line-overloading $\boldsymbol{H}^*$ from P2 to P3} \\ \hline \hline
\end{tabular}
\vspace{-9pt}
\end{table*}

\subsection{Proposed Corrective Cyber-Defense Strategy}

In the absence of cyber-attacks, power systems tend to operate at the cheapest operation point, which is achieved by fully making use of line flow limits, leading to zero cybersecurity margin and thus being dangerous.
In the presence of cyber-attacks, power systems tend to operate in the safest operation point (i.e., an operation point with the largest cybersecurity margin), which can not fully make use of the line flow limits and thus lead to additional defense-induced costs.

Mathematically, the safest point (i.e., the point with the maximum security margin) inside a polytope is referred to as the Euclidean Chebyshev center~\cite{boyd2004convex}, i.e., the center of the largest hyperball that lies inside the polytope.
Thus, the safest operation point $\boldsymbol{G}$ with the maximum cybersecurity margin $r$ can be obtained by solving the linear programming problem~\cite{boyd2004convex}:
\begin{equation}
    \max \quad r
\end{equation}

over~~~~~~~~~~~~~~$G \in \mathbb{R}^G$, $r \in \mathbb{R}^+$

s.t.~~~~~~~~~~~~~~~\eqref{eq_powerbalance},~\eqref{eq_Genlimit},~\eqref{eq_AGb},~\eqref{eq_Ab}
\begin{equation} \label{eq_AGbri}
    \sup_{\| \boldsymbol{w} \|_2 \le r} \boldsymbol{A}_i (\boldsymbol{G} + \boldsymbol{w} ) = \boldsymbol{A}_i \boldsymbol{G} + r \| \boldsymbol{A}_i \|_2 \le b_i \quad \forall i
\end{equation}

In this regard, the cyber-defense benefit of dispatching the operation point is defined as the cybersecurity margin:
\begin{equation} \label{eq_dispatch_BG}
    B^G = r
\end{equation}

The cyber-defense cost of dispatching is defined as the additional defense-induced operation cost compared with the most economical operation point (which is a constant and thus can be omitted):
\begin{equation} \label{eq_dispatch_CG}
    C^G = \boldsymbol{c}^T \boldsymbol{G}
\end{equation}
where $\boldsymbol{c}$ is the generation cost vector.

Similar to P1$'$, the system operators aim to optimize two conflicting objectives.
One objective is to maximize the cybersecurity margin $B^G$.
The other objective is to minimize the additional defense-induced operation point $C^G$.
This yields the following bi-objective linear programming problem:
\begin{equation} \label{eq_objG_P3_}
	\text{P3}': \ \min_{G \in \mathbb{R}^G, r \in \mathbb{R}^+} \ \{ -B^G,~C^G \} \quad \quad \quad \quad
\end{equation}
~~~~~~~~~~~~~~~~~~~~~~s.t.~\eqref{eq_powerbalance},~\eqref{eq_Genlimit},~\eqref{eq_AGb},~\eqref{eq_Ab},~\eqref{eq_AGbri},~\eqref{eq_dispatch_BG},~\eqref{eq_dispatch_CG}

To obtain the Pareto optimal front, we adopt the weighted sum method~\cite{MulObj_weighted_2010}, yielding the following single-objective problem:
\begin{equation} \label{eq_objG_P3}
    \text{P3}: \ \min_{G \in \mathbb{R}^G, r \in \mathbb{R}^+} \  -B^G + \omega^G C^G \quad \quad \quad \quad
\end{equation}
~~~~~~~~~~~~~~~~~~~~~~s.t.~\eqref{eq_powerbalance},~\eqref{eq_Genlimit},~\eqref{eq_AGb},~\eqref{eq_Ab},~\eqref{eq_AGbri},~\eqref{eq_dispatch_BG},~\eqref{eq_dispatch_CG}

\noindent
where $\omega^G$ represents the weight coefficient to balance the cyber-defense cost and benefit.
Similar to P1$'$, the Pareto optimal front of P3$'$ can be obtained by varying $\omega^G$.
Consequently, the Pareto optimal front enables the system operators with a corrective cyber-defense strategy, which indicates how to maximize $B^G$ via optimally dispatching $G$ with respect to any given cyber-defense cost budget.
This leads to a cyber-defense cost-benefit trade-off between the \textit{safest-but-expensive} operation point (i.e., Euclidean the Chebyshev center) and the \textit{cheapest-but-dangerous} operation point in Fig.~\ref{fig_Redis}.

\vspace{-8pt}

\subsection{Two Pareto Optimal Fronts of the Preventive-Corrective Cyber-Defense Strategy}

The proposed preventive strategy in system planning and corrective strategy in system operation form a preventive-corrective cyber-defense strategy, which is illustrated by Table~\ref{table_cyberCostBene} and realized by Algorithm~1.

Physically, the proposed preventive cyber-defense strategy, which is implemented in system planning, provides a $C^{\text{AIR}}-B^{\text{AIR}}$ Pareto optimal front for system operators with the following information:
\begin{enumerate}
    \item what is the optimal cyber-defense benefit (i.e., the minimum FDI attack-induced region volume) with respect to any given cyber-defense cost budget (i.e., the number of measurement protections);
    \item how to optimally allocate the limited measurement protection to achieve the optimal cyber-defense benefit.
\end{enumerate}

After determining the preventive cyber-defense strategy in system planning, the consequent maximum line-overloading $\boldsymbol{H}^*$ is delivered to the corrective cyber-defense strategy.
Similarly, the proposed corrective cyber-defense strategy, which is implemented in system operation, provides a $C^G-B^G$ Pareto optimal front for system operators with the following information:
\begin{enumerate}
    \item what is the optimal cyber-defense benefit (i.e., the maximum cybersecurity margin) with respect to any given cyber-defense cost budget (i.e., the defense-induced generation cost);
    \item how to optimally redispatch the generations to achieve the optimal cyber-defense benefit.
\end{enumerate}

\begin{algorithm}[!t]
    \renewcommand{\algorithmicrequire}{\textbf{Input:}}
	\renewcommand{\algorithmicensure}{\textbf{Output:}}
    \caption{Two Pareto Optimal Fronts of the Preventive-Corrective Cyber-Defense Strategy}
	\begin{algorithmic}[1]
	\small{
        \REQUIRE network topology (shift factor matrix $\boldsymbol{S}$, incidence matrix $\boldsymbol{U_G}$, $\boldsymbol{U_D}$); generator limits $\underline{\boldsymbol{G}}$, $\overline{\boldsymbol{G}}$; line limits $\underline{\boldsymbol{F}}$, $\overline{\boldsymbol{F}}$.
	\ENSURE Preventive cyber-defense strategy: $C^{\text{AIR}}-B^{\text{AIR}}$ Pareto optimal front and corresponding measurement protection strategies.
        \ENSURE Corrective cyber-defense strategy: $C^G-B^G$ Pareto optimal front and corresponding generation redispatch strategies.
    \STATE \textbf{Preventive cyber-defense strategy}:
    \STATE initialize a sufficiently large $\omega^{\text{AIR}}_M$ and solve P2, leading to the smallest $C^{\text{AIR}}_m = 0$ and thus the smallest $B^{\text{AIR}}_m$.
    \STATE initialize a sufficiently small $\omega^{\text{AIR}}_m$ and solve P2, leading to the largest $C^{\text{AIR}}_M$ and thus the largest $B^{\text{AIR}}_M$.
    \STATE adjust $\omega^{\text{AIR}}$ within $[\omega^{\text{AIR}}_m,~\omega^{\text{AIR}}_M]$ using binary search algorithm and solve P2, yielding the $C^{\text{AIR}}-B^{\text{AIR}}$ Pareto optimal front and corresponding measurement protection strategies.
    \STATE \textbf{Information transmission}: 
    \STATE deliver the consequent maximum line-overloading $\boldsymbol{H}^*$ from the preventive stage to the corrective stage.
    \STATE \textbf{Corrective cyber-defense strategy}:
    \STATE determine $\Gamma^{\text{security}}$ according to $\boldsymbol{H}^*$.
    \STATE initialize a sufficiently large $\omega^G_M$ and solve P3, leading to the smallest $C^G_m$ and thus the smallest $B^G_m$.
    \STATE initialize a sufficiently small $\omega^G_m$ and solve P3, leading to the largest $C^G_M$ and thus the largest $B^G_M$.
    \STATE adjust $\omega^G$ within $[\omega^G_m,~\omega^G_M]$ using binary search algorithm and solve P3, yielding the $C^G-B^G$ Pareto optimal front and corresponding generation redispatch strategies.}
	\end{algorithmic}
\end{algorithm}

Mathematically, the proposed preventive cyber-defense strategy P2 focuses on optimizing the shape of the set $\Omega^{\text{AIR}}$, while the proposed corrective cyber-defense strategy P3 focuses on finding an optimal element $\boldsymbol{G}$ inside the set $\Gamma^{\text{security}}$.
Note that $\Gamma^{\text{security}}$ in P3 is determined by $\Omega^{\text{AIR}}$ in P2 via $\boldsymbol{H}^*$.

\section{Case Study}

To verify the effectiveness and the Pareto optimal front of the proposed preventive-corrective cyber-defense strategy, this section conducts case studies on 1) a modified IEEE 14 bus system~\cite{MatPower_TPS2011} with an additional 0.1~pu load at bus 8; and 2) the New England IEEE 39 bus system~\cite{MatPower_TPS2011}.
In addition, the feasibility of online implementation of the corrective cyber-defense strategy in system operation is verified in IEEE 57/118/300 bus systems.
The optimization problems are solved by Yalmip with Gurobi in a personal computer with 16 GB RAM and two 2.30 GHz processors.
The power base is 100~MW.
The FDI cyber-attack ability is set as $\tau = 0.5$~\cite{LRload_2012}, indicating the cyber-attack injection is limited by 50\% of the bus load.

\begin{figure}[!b]
    \centering
    \includegraphics[width=8.8cm]{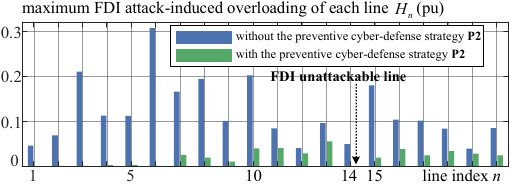}\\
    \vspace{-7pt}
    \caption{Comparison of the maximum FDI attack-induced line overloading in Base Case with and without the preventive cyber-defense strategy P2 in the modified IEEE 14 bus system}
    \label{fig_HeachLine}
\end{figure}

\subsection{Effectiveness of the Proposed Preventive Cyber-Defense Strategy P2 in the Modified IEEE 14 Bus System}

In the modified IEEE 14 bus system,
there are 12 loads (except for bus~1 and bus~7) and 20 lines, whose indices are consistent with~\cite{MatPower_TPS2011}.
The total load is 2.69~pu.
We set a Base Case whose parameters are as follows.
For the preventive stage, the cyber-defense cost budget $\overline{C^{\text{AIR}}} = 15$, indicating that at most fifteen measurement protection can be deployed either in loads or in lines.
The line upper limits $\overline{F_n} = 1$~pu except for $\overline{F_1} = 1.5$~pu.
The cyber-defense cost coefficient $\omega^{\text{AIR}} = 0.15$.
The sufficiently large constants $M = 1$, $N = 2$, and $K = 1$, since the lower bounds obtained from Proposition~3 are 0.9399, 1.8797, and 0.9420, respectively.
For the corrective stage, the generation cost vector $\boldsymbol{c} = [20, 30, 60, 50, 25]^T \$ / \text{pu}$.
The generation lower limit $\underline{\boldsymbol{G}} = 0$ and upper limit $\overline{\boldsymbol{G}}=2$~pu.

In the absence of the proposed measurement protection strategy P2, the original FDI attack-induced region volume (denoted as $-B^{\text{AIR}}_0 = \sum_n H_n / \overline{F_n}$) solved by~\eqref{eq_MaxMinObj} is 2.3894~pu.
By solving P2, six measurement protection are placed at six load measurements, whose bus indices are 2, 3, 4, 8, 9, 14, respectively.
No line measurements are equipped with measurement protection, which verifies Remark~2.
As a result, the maximum FDI attack-induced line overloading $H_n$ of each line is illustrated in Fig.~\ref{fig_HeachLine}.
It is observed that, after implementing P2, $H_n$ of each line significantly decreases.
The FDI attack-induced region volume remarkably decreases by 83\% (from $-B^{\text{AIR}}_0 = 2.3894$~pu to $-B^{\text{AIR}} = 0.4072$~pu).
In addition, since the load at bus~8 is equipped with a measurement protection, the two conditions~\eqref{eq_unattack_d} and~\eqref{eq_unattack_l} in Proposition~2 are satisfied, leading to an FDI unattackable line~14, i.e., the line from bus~7 to bus~8.

In short, the proposed preventive cyber-defense strategy P2 successfully decreases the FDI attack-induced region volume $\Omega^{\text{AIR}}$ by optimally placing the limited measurement protection.

\begin{table*}[!t]
    \centering
    \vspace{-6pt}
    \caption{Consequent Cyber-Defense Costs and Benefits with respect to Different Cost Coefficients in P3 in the modified IEEE 14 bus system}
    \label{table_Reshape}
\begin{tabular}{ccccc}
\hline \hline
\begin{tabular}[c]{@{}c@{}}cyber-defense cost\\ coefficient $\omega^G$\end{tabular} & \begin{tabular}[c]{@{}c@{}}consequent cybersecurity margin\\ $B^G=r$~(pu)\end{tabular} & \begin{tabular}[c]{@{}c@{}}nearest boundaries to
\\ the operation point $\boldsymbol{G}$\\\end{tabular} & \begin{tabular}[c]{@{}c@{}}consequent cyber-defense cost\\ $C^G=\boldsymbol{c}^T \boldsymbol{G}$~(\$)\end{tabular} & \begin{tabular}[c]{@{}c@{}}generation\\ $\boldsymbol{G}$\end{tabular} \\ \hline \hline
0.01 & \begin{tabular}[c]{@{}c@{}} 1.00 (\textbf{1900\%}) \\ \textbf{safest} operation point \\ (\textbf{Chebyshev center}) \end{tabular} & $\overline{F_1}$,~$\overline{F_3}$,~$\overline{F_{10}}$,~$\overline{F_{14}}$,~$\underline{F_{14}}$ & \begin{tabular}[c]{@{}c@{}} 95.81 (\textbf{67\%}) \\ expensive operation point  \end{tabular} & [0.38, 1.49, 0.51, 0.21, 0.10]$^T$ \\ \hline
0.015 & 0.84 (1580\%) & $\overline{F_1}$,~$\overline{F_3}$,~$\overline{F_{10}}$,~$\underline{F_{14}}$ & 82.87 (45\%) & [0.69, 1.40, 0.34, 0.00, 0.26]$^T$ \\ \hline
0.03 & 0.60 (1100\%) & $\overline{F_1}$,~$\overline{F_3}$,~$\underline{F_{14}}$ & 67.20 (17\%) & [1.10, 1.09, 0.00, 0.00, 0.50]$^T$ \\ \hline
0.06 & 0.16 (220\%) & $\overline{F_1}$,~$\underline{F_{14}}$ & 58.49 (2\%) & [1.75, 0.00, 0.00, 0.00, 0.94]$^T$ \\ \hline
0.10 & \begin{tabular}[c]{@{}c@{}} 0.05 (\textbf{0\%}) \\ dangerous operation point \end{tabular} & $\overline{F_1}$ & \begin{tabular}[c]{@{}c@{}} 57.25 (\textbf{0\%}) \\ \textbf{cheapest} operation point \end{tabular} & [2.00, 0.00, 0.00, 0.00, 0.69]$^T$ \\ \hline \hline
\end{tabular}
\vspace{-8pt}
\end{table*}

\vspace{-8pt}

\subsection{Pareto Optimal Front of the Proposed Preventive Cyber-Defense Strategy P2 in the Modified IEEE 14 Bus System}

\begin{figure}[!b]
    \centering
    \includegraphics[width=8.8cm]{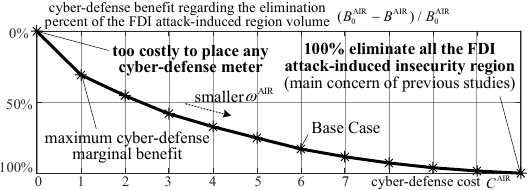}\\
    \vspace{-7pt}
    \caption{Pareto optimal front regarding the cyber-defense benefit and cyber-defense cost of P2 in the modified IEEE 14 bus system}
    \label{fig_ParetoReshape}
\end{figure}

\begin{figure}[!b]
    \centering
    \includegraphics[width=8.8cm]{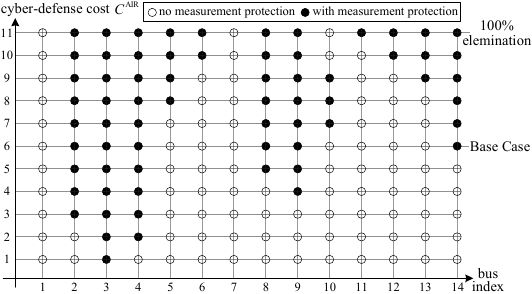}\\
    \vspace{-7pt}
    \caption{Results of the proposed preventive cyber-defense strategy P2 with respect to different cyber-defense costs in the modified IEEE 14 bus system}
    \label{fig_chess}
\end{figure}

The proposed preventive cyber-defense strategy P2 only adopts six measurement protection instead of all fifteen measurement protection,
since P2 needs to concurrently balance both the cyber-defense cost $C^{\text{AIR}}$ and the cyber-defense benefit $B^{\text{AIR}}$ in the objective function~\eqref{eq_objAIR_P2}.
In this regard, by varying the cyber-defense cost coefficient $\omega^{\text{AIR}}$, the Pareto optimal front regarding the trade-off between two cyber-defense objectives, i.e., minimizing the cyber-defense cost and maximizing the cyber-defense benefit in P2, is depicted in Fig.~\ref{fig_ParetoReshape}.
For each cyber-defense cost, the corresponding placement of measurement protection at each load bus is depicted in Fig.~\ref{fig_chess}.

It is observed that, the more expensive the cyber-defense cost (i.e., larger $\omega^{\text{AIR}}$), the less placement the measurement protection (i.e., smaller $C^{\text{AIR}}$).
Specifically, if the cyber-defense cost is negligible (e.g., $\omega^{\text{AIR}} = 0.01$), P2 decides to place eleven measurement protection on the load buses in Fig.~\ref{fig_chess}, leading to $\boldsymbol{H} = \boldsymbol{0}$ and $(B^{\text{AIR}}_0 - B^{\text{AIR}})/B^{\text{AIR}}_0 = 100\%$ in Fig.~\ref{fig_ParetoReshape}.
In other words, the system operators need at least eleven measurement protection to completely (100\%) eliminate the FDI attack-induced region $\Omega^{\text{AIR}}$,
which is the main concern of the previous studies~\cite{bobba2010detecting, TII2021_GraphSensor, TSG2014_GraphSensor, TSG2013_Game_Sensor, TII2015_Greedy_Sensor, TSG2019_Comb_Sensor}.
To avoid wasting cyber-defense resources, the rest four measurement protection do not need to be placed.
By contrast, if the cyber-defense cost is too expensive (e.g., $\omega^{\text{AIR}} = 1$), P2 decides to place no measurement protection. 

In addition, the cyber-defense marginal benefit decreases if the measurement protection increases.
That is, the maximum cyber-defense marginal benefit is achieved when placing only one measurement protection (at the load whose bus index is 3), which decreases the FDI attack-induced region volume by 31\%.

\vspace{-8pt}

\subsection{Effectiveness and Cost-Effectiveness of the Proposed Corrective Cyber-Defense Strategy P3 in the Modified IEEE 14 Bus System}

Due to a limited cyber-defense cost budget $\overline{C^{\text{AIR}}}$ or too large cyber-defense cost coefficient $\omega^{\text{AIR}}$, the FDI attack-induced region $\Omega^{\text{AIR}}$ may not be completely eliminated by the proposed preventive cyber-defense strategy P2, leading to $\boldsymbol{H} \ne \boldsymbol{0}$.
Then, the proposed corrective cyber-defense strategy P3 is implemented after P2.

\begin{figure}[!t]
    \centering
    \includegraphics[width=8.8cm]{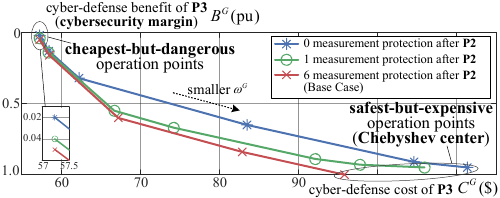}\\
    \vspace{-7pt}
    \caption{Pareto optimal front regarding the cyber-defense benefit and cost of P3 after different P2 decisions in the modified IEEE 14 bus system}
    \vspace{-9pt}
    \label{fig_ParetoRedis}
\end{figure}

By varying the cyber-defense cost coefficient $\omega^G$, the consequent cyber-defense cost (i.e., the operation cost $C^G=\boldsymbol{c}^T \boldsymbol{G}$) and cyber-defense benefit (i.e., the cybersecurity margin $B^G = r$) of P3 (after implementing Base Case in P2) are listed in Table~\ref{table_Reshape}.
It is observed that a larger $\omega^G$ leads to a smaller operation cost but also a smaller cybersecurity margin.
Specifically, if $\omega^G$ is large enough (e.g., $\omega^G = 0.1$), the consequent generation is $\boldsymbol{G}_0 = [2, 0, 0, 0, 0.69]^T$~pu, leading to the cheapest operation cost $C^{G_0} = \boldsymbol{c}^T \boldsymbol{G}_0 = 57.25 \$$.
Note that the total power generation $\boldsymbol{1}^T \boldsymbol{G}$ is 2.69~pu and equals the total power consumption.
The minimum distance from the generation point $\boldsymbol{G}_0$ to the boundaries is $q_1 = 0.05$~pu, i.e., the distance to $\overline{F_1}$.
That is, the cybersecurity margin $B^{G_0} = 0.05$~pu.
This leads to a \textit{cheapest-but-dangerous} operation point.

By contrast, if $\omega^G$ is small enough (e.g., $\omega^G = 0.01$), the consequent generation $\boldsymbol{G} = [0.38, 1.49, 0.51, 0.21, 0.10]^T$~pu.
Note that the total power generation $\boldsymbol{1}^T \boldsymbol{G}$ remains 2.69~pu and ensures the power flow balance.
The operation cost is 95.81\$, which increases by 67\% compared to the cheapest operation cost $C^{G_0}$.
However, the cybersecurity margin is $B^G = r = 1$~pu, which increases by 1900\% compared to $B^{G_0}=0.05$~pu.
In other words, the proposed corrective cyber-defense strategy P3 manages to move the cheapest-but-dangerous operation point to a \textit{safest-but-expensive} operation point. 
In addition, the cyber-defense marginal benefit decreases if the cyber-defense cost increases.
The maximum cyber-defense marginal benefit is achieved if $\boldsymbol{G} = [1.75, 0, 0, 0, 0.94]$, leading to 2\% additional defense-induced operation cost but 220\% increment of cybersecurity margin.

As shown in Fig.~\ref{fig_ParetoRedis}, the Pareto optimal front regarding the cyber-defense cost-benefit trade-off of P3 is influenced by the different P2 decisions.
Although more measurement protection from P2 only sightly increase the P3 cyber-defense benefits, the P3 cyber-defense costs are significantly decreased when the safest-but-expensive operation points are chosen.

\vspace{-8pt}

\subsection{Preventive-Corrective Cyber-Defense Strategy in the New England IEEE 39 Bus System}

\begin{figure}[!b]
    \centering
    \includegraphics[width=8.8cm]{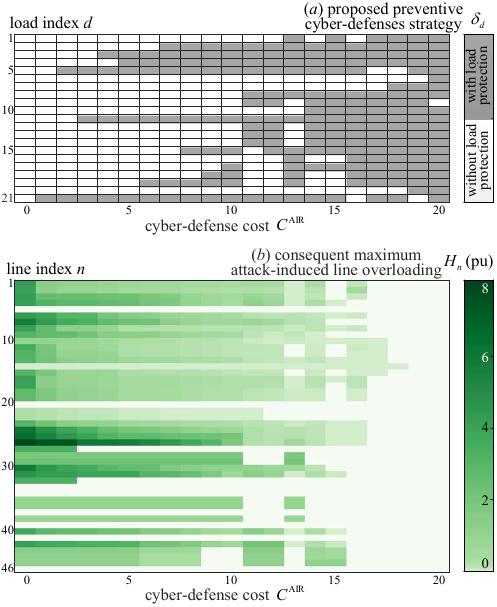}\\
    \vspace{-7pt}
    \caption{Preventive cyber-defense strategy P2 for the New England IEEE 39 bus system with respect to any given cyber-defense cost. (\textit{a}) proposed measurement protection. (\textit{b}) consequent maximum attack-induced line overloading.}
    \vspace{-11pt}
    \label{fig_39BusHeatMap}
\end{figure}

The New England IEEE 39 bus system has 21 loads, 46 lines, and 10 generators, whose indices are consistent with~\cite{MatPower_TPS2011}.
The total load is 62.54~pu.
For the preventive stage, the line upper limits $\overline{F_n}$ is 15\% of total loads except for the lines connecting generators.
For the corrective stage, the generation cost vector is randomly set as $c=[84, 92, 21,	93,	67,	18,	35,	59,	97,	97]^T$ \$/pu. The generation lower limit $\underline{\boldsymbol{G}} = 0$ and upper limit $\overline{\boldsymbol{G}}$ equals the total loads.

According to the proposed preventive cyber-defense strategy P2, the load measurement protection with respect to different cyber-defense cost $C^{\text{AIR}}$ is depicted in Fig.~\ref{fig_39BusHeatMap}(\textit{a}), leading to the maximum attack-induced line overloading $H_n$ in Fig.~\ref{fig_39BusHeatMap}(\textit{b}).
It is observed that the proposed preventive cyber-defense strategy manages to gradually decrease the FDI attack-induced region volume by optimally placing any given measurement protection budget.
In addition, several FDI unattackable lines (e.g., line 5, 20, 33, 34, 37, 39, 41, 46) are identified by satisfying the Proposition~2.

\begin{figure}[!t]
    \centering
    \includegraphics[width=8.8cm]{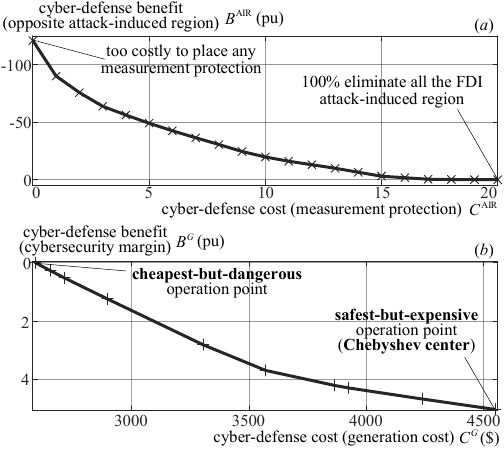}\\
    \vspace{-7pt}
    \caption{Cyber-Defense cost-benefit Pareto optimal fronts in the New England IEEE 39 bus system regarding (\textit{a}) the proposed preventive cyber-defense strategy P2; and (\textit{b}) the proposed corrective cyber-defense strategy P3}
    \vspace{-8pt}
    \label{fig_39BusPareto}
\end{figure}

Furthermore, the $C^{\text{AIR}}-B^{\text{AIR}}$ Pareto optimal front of the proposed preventive cyber-defense strategy P2 is depicted in Fig.~\ref{fig_39BusPareto}(\textit{a}), which assists power system operators to optimally allocate any given measurement protection in system planning stage such that the FDI attack-induced region is minimized.
Assuming $\boldsymbol{H}^* = \boldsymbol{0}$, the $C^G-B^G$ Pareto optimal front of the proposed corrective cyber-defense strategy P3 is depicted in Fig.~\ref{fig_39BusPareto}(\textit{b}), which helps power system operators to optimally redispatch any cyber-defense cost budget such that the cybersecurity margin is maximized.

\subsection{Computational Performance in IEEE 57/118/300 Bus Systems}

\begin{table}[!b]
    \centering
    \vspace{-6pt}
    \caption{Average Solution Time of the Proposed Corrective Cyber-Defense Strategy P3}
    \label{table_AvgTime}
\begin{tabular}{cccccc}
\hline
test IEEE system & \begin{tabular}[c]{@{}c@{}} 14 \\bus\end{tabular}   & \begin{tabular}[c]{@{}c@{}} 39 \\bus\end{tabular}   & \begin{tabular}[c]{@{}c@{}} 57 \\bus\end{tabular}   & \begin{tabular}[c]{@{}c@{}} 118 \\bus\end{tabular}  & \begin{tabular}[c]{@{}c@{}} 300 \\bus\end{tabular}   \\ \hline
\begin{tabular}[c]{@{}c@{}}average\\ solution time (s)\end{tabular} & 1.05 & 1.32 & 1.66 & 4.27 & 10.83 \\ \hline
\end{tabular}
\end{table}

Compared with the preventive cyber-defense strategy P2 that is implemented in system planning, the corrective cyber-defense strategy P3 is implemented in the system operation stage and thus requires higher computational performance.
Note that P3 is a linear programming problem, which can be effectively solved using off-the-shelf solvers.
By varying the weights $\omega^G$ in P3, we test the average solution time of P3 (by repeating 100 times) in different systems, including the modified IEEE 14 bus system, the New England IEEE 39 bus system, IEEE 57 bus system, IEEE 118 bus system, and IEEE 300 bus system.
As shown in Table~\ref{table_AvgTime}, the solution time of P3 in these test systems is significantly lower than the time period of the optimal power flow, indicating the feasibility of online implementation in system operation.

\section{Conclusion}

Considering any given cyber-defense resource, this paper proposes a cost-effective preventive-corrective cyber-defense strategy.
First, this paper proposes a preventive cyber-defense strategy that minimizes the volume of the FDI attack-induced region with respect to any given measurement protection.
Then, given the consequent preventive security region, this paper proposes a corrective cyber-defense strategy to achieve a trade-off between maximizing the cybersecurity margin and minimizing the additional defense-induced operation cost.

For the proposed preventive cyber-defense strategy in the system planning stage, the case study indicates that it optimally shapes the FDI attack-induced region with respect to any given cyber-defense cost budget and coefficient.
The optimal measurement protection leads to an FDI unattackable line, indicating the locally rather than globally prevented FDI cyber-attacks.
Moreover, the smallest subset of measurement protection to 100\% eliminate the FDI attack-induced region is identified, which is the main concern of previous studies~\cite{bobba2010detecting, TII2021_GraphSensor, TSG2014_GraphSensor, TSG2013_Game_Sensor, TII2015_Greedy_Sensor, TSG2019_Comb_Sensor}.
In addition, it is found that, using only one measurement protection, the FDI attack-induced region volume is decreased by 31\% in the modified IEEE 14 bus system, indicating the maximum cyber-defense marginal benefit.

For the proposed corrective cyber-defense strategy in the system operation stage, the case study indicates that it successfully balances the cybersecurity margin and the additional defense-induced operation cost within the preventive security region.
The trade-off between the safest-but-expensive (i.e., the Chebyshev center) and the cheapest-but-dangerous operation point are addressed.
In addition, it is observed that, in the modified IEEE 14 bus system, the maximum cyber-defense marginal benefit leads to 220\% increment of cybersecurity margin with only 2\% additional defense-induced operation cost.

\appendices

\section{Proof of Proposition~2} \label{ap_unattack}

\begin{figure}[!t]
    \centering
    \vspace{-7pt}
    \includegraphics[width=6.7cm]{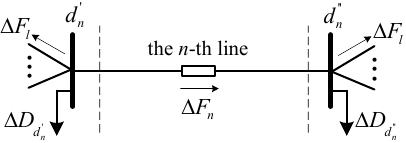}\\
    \caption{The $n$-th line with terminal bus $d_n'$ and $d_n''$}
    \vspace{-8pt}
    \label{fig_unattack}
\end{figure}

\noindent
\textit{Proof.}
As shown in Fig.~\ref{fig_unattack}, $d_n'$ and $d_n''$ are the two terminal buses of the $n$-th line.
Take bus~$d_n'$ as an example, the total FDI cyber-attack injections (except for $\Delta F_n$) at bus~$d_n'$ is
\begin{equation}
    \Delta P_{d_n'} = \Delta D_{d_n'} + \sum_{l \in \mathcal{N}_{d_n'}} \Delta F_l
\end{equation}
where $\mathcal{N}_{d_n'}$ denotes the set of all lines interconnected to the bus $d_n'$ except for the $n$-th line. 

To satisfy one of the stealthy requirements, i.e., the power flow equation $\Delta \boldsymbol{F} = \boldsymbol{S} \boldsymbol{U_D} \Delta \boldsymbol{D}$ in~\eqref{eq_FDI_FmLimit},
the FDI attack net injection at bus~$d_n'$ should be zero, yielding
\begin{equation}
    0 = \Delta F_n + \Delta P_{d_n'}
\end{equation}

If $\delta_{d_n'} = 0$ in~\eqref{eq_unattack_d} and $\sum_{l \in \mathcal{N}_{d_n'}} \varepsilon_l = 0$ in~\eqref{eq_unattack_l} hold, both the $\Delta D_{d_n'}$ and $\Delta F_l, l \in \mathcal{N}_{d_n'}$ are disabled.
Hence, we have $\Delta P_{d_n'} = 0$ and thus $\Delta F_n = 0$, leading to~\eqref{eq_Hn0Ln0}.~~~~~~~~~~~~~~~~~~$\square$

\section{Proof of Proposition~3} \label{ap_bigM}

\noindent
\textit{Proof.}
According to~\eqref{eq_FDI_FmLimit_Dn} and~\eqref{eq_FDI_DdLimit_Dn}, we have
\begin{alignat}{1}
    \boldsymbol{S} \boldsymbol{U_D} \Delta \boldsymbol{D} &\le | \boldsymbol{S} \boldsymbol{U_D} \Delta \boldsymbol{D} | \le | \boldsymbol{S} \boldsymbol{U_D} | \ |  \Delta \boldsymbol{D}  | \notag \\
    &\le | \boldsymbol{S} \boldsymbol{U_D} | \ ( \boldsymbol{\delta} \circ \boldsymbol{\tau} \circ \boldsymbol{D} ) \le | \boldsymbol{S} \boldsymbol{U_D} | \ ( \boldsymbol{\tau} \circ \boldsymbol{D} )
\end{alignat}

Thus, a lower bound of $M$ in~\eqref{eq_betagroup} is the maximum element of the vector $| \boldsymbol{S} \boldsymbol{U_D} | \ ( \boldsymbol{\tau} \circ \boldsymbol{D} )$, yielding~\eqref{eq_bigM_M}.

Similarly, according to~\eqref{eq_betagroup}, we have
\begin{equation}
    M \pm \boldsymbol{S} \boldsymbol{U_D} \Delta \boldsymbol{D} \le M + | \boldsymbol{S} \boldsymbol{U_D} | \ ( \boldsymbol{\tau} \circ \boldsymbol{D} )
\end{equation}

Thus, a lower bound of $N$ in~\eqref{eq_betagroup} is the maximum element of the vector $M + | \boldsymbol{S} \boldsymbol{U_D} | \ ( \boldsymbol{\tau} \circ \boldsymbol{D} )$, yielding~\eqref{eq_bigM_N}.

According to~\eqref{eq_alphagroup} and~\eqref{eq_FDI_DdLimit_Dn}, we have
\begin{equation}
    \boldsymbol{\delta} \circ \boldsymbol{\tau} \circ \boldsymbol{D} \pm \Delta \boldsymbol{D} \le  \boldsymbol{\tau} \circ \boldsymbol{D}  +  \boldsymbol{\tau} \circ \boldsymbol{D} 
\end{equation}

Thus, a lower bound of $K$ in~\eqref{eq_alphagroup} is the maximum element of the vector $2 \boldsymbol{\tau} \circ \boldsymbol{D} $, yielding~\eqref{eq_bigM_K}.~~~~~~~~~~~~~~~~~~~~~~~~~~~~~~~~~~~$\square$


\ifCLASSOPTIONcaptionsoff
  \newpage
\fi

\bibliographystyle{IEEEtran}
\bibliography{IEEEabrv, myBIB}


%



\end{document}